\pdfoutput=1

\documentclass{jfm}

\usepackage{graphicx, amsmath} 
\usepackage[usenames]{color}
\usepackage{marvosym}
\usepackage{amssymb}

\let\oldepsilon\epsilon
\let\epsilon\varepsilon
\let\varepsilon\oldepsilon

\newcommand{\dx}{\partial_x}

\newcommand{\dt}{\partial_t}

\newcommand{\Tsat}{T_{\rm sat}}
\newcommand{\Lsat}{L_{\rm sat}}
\newcommand{\qsat}{q_{\rm sat}}

\title[Transport relaxation time and length scales in turbulent suspensions]{Transport relaxation time and length scales in turbulent suspensions} 
\author[P. Claudin, F. Charru and B. Andreotti]
{P\ls H\ls I\ls L\ls I\ls P\ls P\ls E \ns C\ls L\ls A\ls U\ls D\ls I\ls N, \ns F\ls R\ls A\ls N\ls \c{C}\ls O\ls I\ls S \ns C\ls H\ls A\ls R\ls R\ls U$^\star$ \and \ns B\ls R\ls U\ls N\ls O\ns A\ls N\ls D\ls R\ls E\ls O\ls T\ls T\ls I \ns} 
\affiliation{
Laboratoire de Physique et M\'ecanique des Milieux H\'et\'erog\`enes,\\
(PMMH UMR 7636 ESPCI - CNRS - Univ. Paris Diderot - Univ. P. M. Curie)\\
10 rue Vauquelin, 75005 Paris, France.\\ \vspace*{0.2cm}
$^\star$Institut de M\'ecanique des Fluides de Toulouse -- CNRS - Universit\'e de Toulouse,\\
31400 Toulouse, France.
}
\pubyear{???}
\volume{???}
\pagerange{??--??}
\date{\today}
\begin{document}
\maketitle

\begin{abstract}
We show that in a turbulent flow transporting suspended sediment, the unsaturated sediment flux $q(x,t)$ can be described by a first-order relaxation equation. From a mode analysis of the advection-diffusion equation for the particle concentration, the relaxation length and time scales of the dominant mode are shown to be the deposition length $H U/V_{\rm fall}$ and deposition time $H/V_{\rm fall}$, where $H$ is the flow depth, $U$ the mean flow velocity and $V_{\rm fall}$ the sediment settling velocity. This result is expected to be particularly relevant for the case of sediment transport in slowly varying flows, where the flux is never far from saturation. Predictions are shown to be in quantitative agreement with flume experiments, for both net erosion and net deposition situations.
\end{abstract} 

\section{Introduction}   \label{intro}

Suspension is an important mode for the transport of sediments by fluid flows. It occurs when the falling velocity of the particles is smaller than the turbulent velocity fluctuations, so that particles can remain suspended for a long time, trapped by turbulent eddies, before they eventually fall back on the bed due to gravity. In nature, one observes suspension in large rivers, \textit{i.e.} in their downstream part, where large amount of fine particles have been collected from the catchment basin. Rivers that ordinarily present bed-load transport (the moving particles remain close to the bed) can also experience suspension (the particles are present over the whole flow depth) when the water discharge is unusually large, e.g. during flood events.

\begin{figure} 
\begin{center}
\setlength{\unitlength}{1mm}
\begin{picture}(100, 60)(0, 0)
\put(0, 0){\includegraphics[width=10cm]{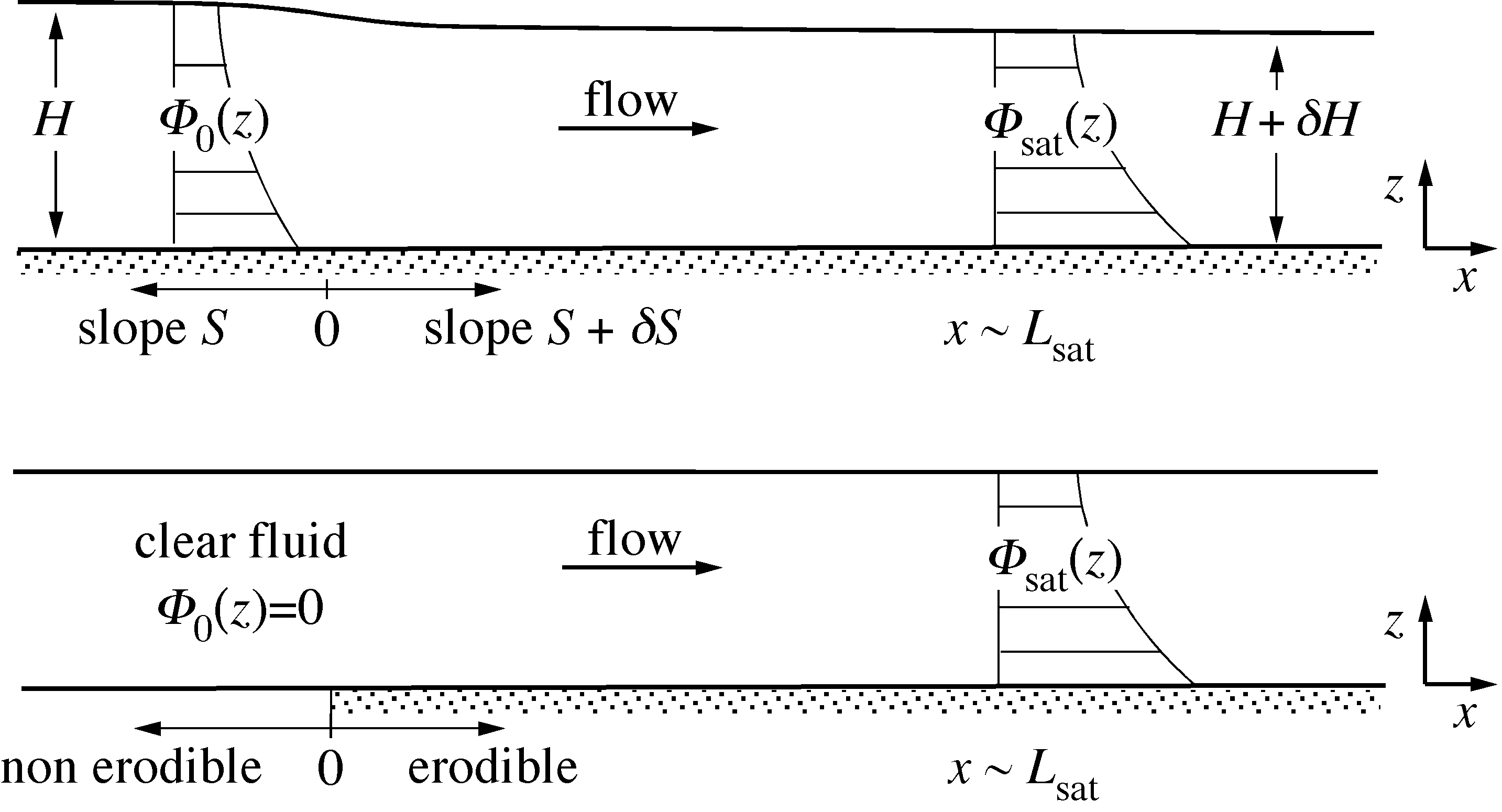}}
\put(-7, 55){(a)}
\put(-7, 22){(b)}
\end{picture}
\end{center}
\caption{Sketches of the situations studied in \S\ref{sec:applic}: (a), flow over a slope change; (b), flow over the passage from non erodible to erodible bed. $H$ and $S$ are the river depth and bottom slope for $x<0$, and $\Phi_0(z)$ is the corresponding saturated concentration profile; $\delta H$ and $\delta S$ are the variations in $H$ and $S$ downstream of the change at $x=0$, and $\Phi_{sat}(z)$ is the new saturated concentration profile reached at the distance $x \approx L_{sat}$. }
\label{fig:sketches}
\end{figure}

Vertical concentration profiles and overall sediment fluxes are among the major issues -- see the pioneering works of Rouse (1936) or Vanoni (1946). From the point of view of hydraulic engineering, the problem is satisfactorily solved for rivers in a steady state, although some questions are still open, such as particle trapping by turbulent eddies, or the structure of the flow near the bottom where the concentration is large (Nielsen 1992; Nezu 2005). However, the response of the sediment flux to temporal or spatial changes of the flow is largely unknown. Such changes may be induced, for instance, by long gravity waves, or a sudden increase of the flow rate, or variations of the river slope or geometry. Two typical problems of relaxation downstream a change in the flow conditions are depicted in Figure~\ref{fig:sketches}, which will be studied in \S\ref{sec:applic}: that of a small change of the slope of the bottom (Fig.~\ref{fig:sketches}a), and that of a change of the bottom conditions, from non-erodible to erodible (Fig.~\ref{fig:sketches}b). The suspended sediment response is expected to have a strong effect on the dynamics of the erodible bottom, especially on the formation of dunes or bars, or, at larger scale, on the development of meanders (Seminara 2006). Specific relaxation problems have been investigated by numerical integration of the Reynolds-averaged Navier-Stokes equations, using mixing length or $k-\epsilon$ turbulence models (Hjelmfelt \& Lenau 1970; Jobson \& Sayre 1970; Apmann and Rumer 1970; van Rijn 1986a; Celik \& Rodi 1988; Ouillon \& Le Guennec 1996).

In the case where bed-load is dominant, or in the aeolian situation (saltation), it has been shown that the evolution of the sediment flux $q$ can be accounted for by a relaxation equation of the form
\begin{equation}
\Tsat \dt q + \Lsat \dx q = \qsat - q,
\label{eq:relax}
\end{equation}
where $\qsat$ is the saturated flux, $\Tsat$ and $\Lsat$ are the relaxation time and length scales -- \textit{i.e.} the time and length over which the flux relaxes toward saturation. This saturation corresponds to the homogeneous and steady state for which sediment transport is constant both in space and time for given flow conditions. Such a first order equation was first introduced in the aeolian context, as the simplest equation describing relaxation effects (Sauermann, Kroy \& Herrmann 2001; Andreotti, Claudin \& Douady 2002; Kroy, Sauermann \& Herrmann 2002; Andreotti 2004). It was then shown that, for water flows, this equation can be derived from analysis of the erosion and deposition rates, the relaxation scales being there related to particle deposition (and not particle inertia) (Charru 2006; Lajeunesse, Malverti \& Charru 2010). The importance of the relaxation length $\Lsat$ appeared to be crucial in particular for stability analyses of the erodible bottom, for selection of the ripple wavelength (Fourri\`ere, Claudin \& Andreotti 2010). The importance of relaxation phenomena for suspensions in turbulent flow is well-known in the context of hydraulic engineering (Yalin \& Finlaysen 1973; van Rijn 1986b; Celik \& Rodi 1988; Ouillon \& Le Guennec 1996). This importance has also been recognized in the context of geomorphology; in particular, Lague \& Davy (2009) proposed a deposition length of sediment as the relevant transport length. However, a derivation of a relaxation equation of the form (\ref{eq:relax}), from firm hydrodynamic grounds, is still lacking.

In this paper, we discuss the conditions under which an equation of the form (\ref{eq:relax}) can be derived for turbulent flows, when suspension is the dominant mode of transport, with particular emphasis on the identification of the saturation length and time scales. The article is organised as follows. In the next Section, we present the flow models and saturation conditions. In \S\ref{sec:NonEqFlows} we perform a mode analysis of the advection-diffusion equation for the particle concentration applied to unsaturated cases, and then identify the saturation length and time scales of the flux. The relevance of this approach is illustrated in \S\ref{sec:applic} by treating few examples: (i) the effect on the sediment flux of a change in the river slope; (ii) the change from a fixed to an erodible bed and (iii) the deposition of sediments from a source near the free surface. For the last two situations, the predictions of the model are tested against experimental data from the literature.

\section{Flow models}   \label{FlowModel}

\subsection{Logarithmic flow model}   \label{sec:log_model}

We consider the free-surface, turbulent flow of a fluid layer of thickness $H$ over an erodible bed. For the sake of simplicity, we restrict the discussion to flows invariant in the spanwise direction, \textit{i.e.} two-dimensional, with streamwise coordinate $x$ and upwards transverse coordinate $z$. Measurements have shown that the profile of the streamwise velocity is close to the logarithmic law
\begin{equation}
u_x(z) = \frac{u_*}{\kappa} \, \ln \left(\frac{z + z_0}{z_0}\right)\;,
\label{eq:log_ux}
\end{equation}
where $u_*$ is the friction velocity, $z_0$ is the hydrodynamical bed roughness, and $\kappa = 0.4$ is the von K\'arm\'an coefficient. For steady flow where the shear stress is balanced by the streamwise component of gravity, the shear stress increases linearly from zero at the free surface to $\tau_b = \rho u_*^2$ at the bottom, so that the logarithmic velocity profile (\ref{eq:log_ux}) corresponds to a parabolic eddy viscosity $\nu_t$ (Nezu \& Rodi 1986), given by
\begin{equation}
\frac{\nu_t}{u_* H} = \kappa\;\left(\frac{z + z_0}{H}\right)\; \left( 1-\frac{z}{H} \right).
\label{eq:nu_t}
\end{equation}
From (\ref{eq:log_ux}), the depth-averaged velocity $U$ is given by
\begin{equation}
\lambda \equiv \frac{U}{u_*} = 
\frac{1}{\kappa} \left[ \ln\left(\frac{H}{z_0}\right) - 1 \right],
\label{eq:lambda}
\end{equation}
with typical value $\lambda = 10$, corresponding to $z_0/H \approx 0.01$ (Raudkivi 1998).

We assume that the sediment concentration $\phi$ is governed by the advection-diffusion equation
\begin{equation}
\frac{\partial \phi}{\partial t} + u_x \frac{\partial \phi}{\partial x} = 
\frac{\partial}{\partial x} \left(D \frac{\partial \phi}{\partial x} \right) + \frac{\partial}{\partial z} \left(D \frac{\partial \phi}{\partial z} + 
\phi V_{\rm fall} \right),
\label{eq:concentration}
\end{equation}
where $D$ is the particle eddy diffusivity and $V_{\rm fall}$ the settling velocity. Measurements have shown that $D(z)$ is reasonably parabolic and proportional to the eddy viscosity (\ref{eq:nu_t}), with turbulent Schmidt number
\begin{equation}
{\rm Sc} = \frac{\nu_t}{D}
\label{eq:Sc}
\end{equation}
in the range $0.5$--$1$ (Coleman 1970; Celik \& Rodi 1988; Nielsen 1992). The settling velocity $V_{\rm fall}$ is taken uniform, and, when needed for comparison with experiments, equal to that of a single particle in quiescent fluid. Note that the above modelling ignores inertial effects on particle motion, in particular their ejection from the core of vortices and their clustering (Bec et al. 2007; Hunt et al. 2007). We also limit the discussion to dilute suspensions, \textit{i.e.} small volumic particle concentration $\phi$, for which there is no significant feedback of the particles on transport.

Solving (\ref{eq:concentration}) requires two boundary conditions, one at the free surface and one on the sedimentary bed. At the free surface, the net vertical flux vanishes, giving
\begin{equation}
D \frac{\partial \phi}{\partial z} + \phi\,V_{\rm fall} = 0
\qquad{\rm at} \quad z=H.
\label{eq:bc@z=H}
\end{equation}
At the bottom, just above the bedload layer where particles mainly roll and slide on each other, the diffusive flux is equal to the erosion flux $\varphi_\uparrow$, \textit{i.e.} the volume of particles  entrained in suspension per unit time and bed area (Parker 1978; van Rijn 1986a):
\begin{equation}
- D \frac{\partial \phi}{\partial z} = \varphi_\uparrow
\qquad{\rm at} \quad z=0.
\label{eq:bc@z=0}
\end{equation}
The erosion rate, or `pickup function', is generically an increasing function of the basal shear stress above a threshold. Its functional form is determined phenomenologically from experiments and depends on the nature of the bed -- e.g. whether it is consolidated/cohesive or not, composed of grains or containing clay, etc. (Shields 1936; Einstein 1950; Engelund 1970; van Rijn 1984b; Hanson \& Simon 2001; Briaud \& al. 2001; Bonelli et al. 2007).

Two remarks have to be made here. First, the bottom condition (\ref{eq:bc@z=0}) applies for steady and homogeneous as well as unsteady or heterogeneous flows. In the latter case, the erosion flux may be different from the deposition flux, so that the net flux is nonzero, which may lead to variations of the bed topography (not necessarily, as in the experiments to be discussed later). Possible variations in the bed topography will be ignored here. Second remark, the boundary condition (\ref{eq:bc@z=0}) corresponds to a bed allowing unlimited sediment supply. For more general situations (e.g. fixed bed), slightly different boundary conditions have been proposed, see Celik \& Rodi (1988), which however requires an empirical constant or reference concentration near the bottom to be given, which varies along the channel. Finally, assuming that particles have the same mean velocity as the fluid, $u_x$, the flux of suspended particles, per unit length in the spanwise direction, is given by
\begin{equation}
q =  \int_0^H \phi u_x {\rm d}z.
\label{eq:defq}
\end{equation}

\begin{figure}
\begin{center}
\setlength{\unitlength}{1mm}
\begin{picture}(120, 52)(0, 46)
\put(2, 40){\includegraphics[width=12cm]{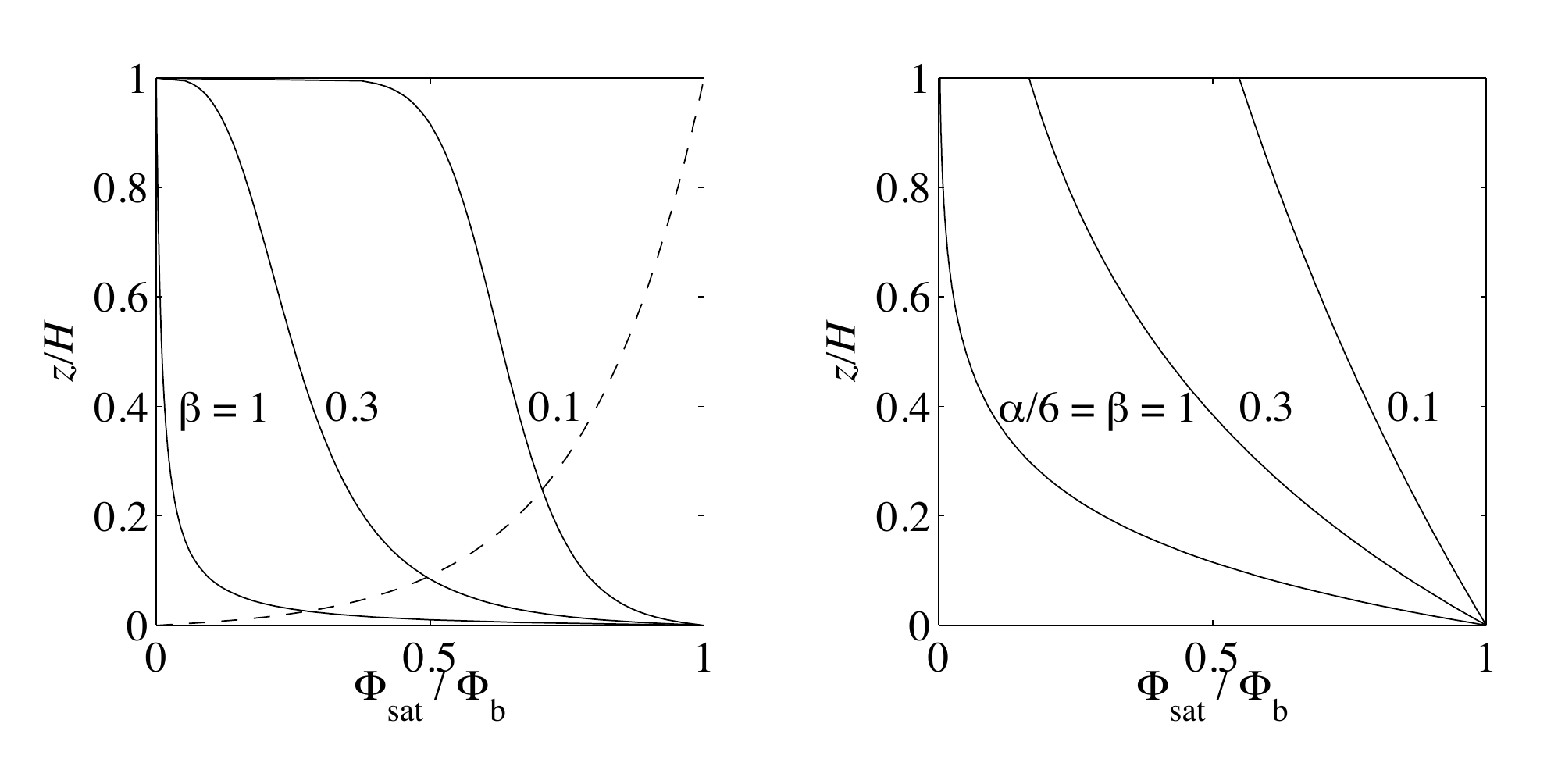}}
\put(3, 91){(a)}
\put(63, 91){(b)}
\end{picture}
\end{center}
\caption{(a) Logarithmic flow model: normalized velocity profile (dashed line) and normalized concentration profiles (\ref{eq:Phi_Rouse}) for three values of $\beta$ (solid lines); (b) plug flow model: normalized concentration profiles (\ref{eq:Phi}) for the corresponding values of $\alpha = 6 \beta$.}
\label{fig:two_models}
\end{figure}

The concentration equation (\ref{eq:concentration}) with the boundary condition (\ref{eq:bc@z=H}) admit a steady and homogeneous solution corresponding to the balance of the settling and diffusive fluxes,
\begin{equation}
\Phi_{\rm sat}(z) = \Phi_{\rm b} \, 
 \left( \frac{1 - z/H}{1 + z/z_0} \right)^{\dfrac{\beta}{1+z_0/H}} ,
\label{eq:Phi_Rouse}
\end{equation}
where $\beta$, known as the Rouse number, is defined as $\beta = ({\rm Sc} \, V_{\rm fall})/(\kappa \, u_*)$ and the bottom concentration $\Phi_{\rm b} = \Phi_{\rm sat}(0)$ is determined from the condition (\ref{eq:bc@z=0}) as
\begin{equation}
\Phi_{\rm b} = \frac{\varphi_\uparrow(\tau_b)}{V_{\rm fall}}.
\label{eq:phi_0}
\end{equation}
Note that (\ref{eq:Phi_Rouse}) differs slightly for the classical expression of the Rouse profile (Nielsen 1992) because the location where the velocity (\ref{eq:log_ux}) vanishes and where the bottom boundary condition (\ref{eq:bc@z=0}) applies has been chosen to be $z=0$ instead of $z=z_0$. Suspension typically occurs when $V_{\rm fall} < 0.8\,u_*$ (Freds{\o}e \& Daigaard 1992), which corresponds to $\beta < 4/3$ with ${\rm Sc} = 2/3$. Figure \ref{fig:two_models}a displays the velocity profile (\ref{eq:log_ux}), normalized by $u_x(H)$, and the concentration profile (\ref{eq:Phi_Rouse}), normalized by $\Phi_{\rm b}$, for three typical values of $\beta$.

\subsection{A simplified plug flow model}   \label{sec:plug_model}

In order to get analytical results, a simplified plug flow model will be used in the following, which appears to provide accurate results as long as assessment of the relaxation equation (\ref{eq:relax}) is pursued. This model corresponds to uniform flow velocity $u_x(z) = U$ and friction velocity $u_* = U / \lambda$, where $\lambda$ is the same constant as in the previous section. Such a plug flow model is of course a rough description, and the velocity profile actually does not switch from zero on the bed to its average value on a vanishing vertical distance, leading to an infinite shear. This problem is not present in the logarithmic model, which is more realistic from this point of view. However, as shown below, these two models do not differ much as far as the relaxation modes are concerned, which means that what occurs very close to the bed is not very important for the present purpose. Accordingly, a uniform particle diffusivity $D_0$ will be taken in the concentration equation (\ref{eq:concentration}) and boundary conditions (\ref{eq:bc@z=H}-\ref{eq:bc@z=0}), equal to the average of the parabolic distribution given by (\ref{eq:nu_t}) and (\ref{eq:Sc}). Up to a small correction of order $z_0/H$, the diffusivity $D_0$ is given by
\begin{equation}
\frac{D_0}{u_* H} = \frac{\kappa}{6 {\rm Sc}} \equiv \mathcal{K}.
\end{equation}

With this uniform diffusivity $D_0$, the advection-diffusion equation (\ref{eq:concentration}) admits the steady and homogeneous solution
\begin{equation}
\Phi_{\rm sat} = \Phi_{\rm b} \, \exp \left( -\alpha \frac{z}{H}\right ) 
\quad {\rm with} \quad 
\alpha \equiv \frac{V_{\rm fall} H}{D_0} = 6 \beta,
\label{eq:Phi}
\end{equation}
which also satisfies the boundary condition (\ref{eq:bc@z=H}) at the free surface. The boundary condition at the bed (\ref{eq:bc@z=0}) determines the bed concentration (\ref{eq:phi_0}). Figure \ref{fig:two_models}b displays the concentration profile (\ref{eq:Phi}), normalized by $\Phi_{\rm b}$, for three typical values of $\alpha = 6\beta$. It can be seen that for the same value of the Rouse number, the plug flow model predicts sediment concentration slightly larger than that of the logarithmic flow model. Small values of $\alpha$ correspond to strong suspensions, i.e. situations for which the sediment is distributed almost uniformly over the whole depth of the flow. This is achieved when the settling velocity is small (very fine particles) or when the diffusivity is large (large flow velocity). Finally, the saturated particle flux per unit width $\qsat$, normalized by the water flux $U H$, is given by
\begin{equation}
\frac{\qsat}{U H} = \frac{1}{U H} \, \int_0^H \Phi_{\rm sat} U {\rm d}z = 
\frac{1 - {\rm e}^{-\alpha}}{\alpha} \, \Phi_{\rm b}.
\label{qsatPlugFlow}
\end{equation}
For small $\alpha$, this dimensionless flux tends to $\Phi_{\rm b}$, as expected.

\section{Non-homogeneous and unsteady flows} \label{sec:NonEqFlows}

In this section, we successively consider a spatial evolution problem (\S \ref{sec:Lsat}) and a temporal evolution problem (\S \ref{sec:Tsat}). These problems are solved using a mode analysis, \textit{i.e.} the departure of the concentation field from the saturated distribution $\Phi_{\rm sat}(z)$ is decomposed as a sum of terms of the form $c(x,t) f(z)$. It is shown that, for the spatial problem, there exists a discrete set of amplitudes $c_n(x) \propto e^{-x/L_n}$, and for the temporal problem, there exists a similar set of amplitudes $c_n(t) \propto e^{-t/T_n}$. Then the sediment flux is shown to be dominated by the mode with the largest length or time, $L_1$ or $T_1$. This result demonstrates that for large scale problems, the relaxation equation (\ref{eq:relax}) retains the most important features of unsaturated sediment transport, with relaxation scales $\Lsat$ and $\Tsat$ equal to the largest scales arising from the mode analysis.

The analytical calculations presented below use the plug flow model because of its simplicity, in the spirit of the work of Mei (1979). Calculations for the logarithmic model are not reported in detail, but the corresponding results are plotted for comparison in some of the figures.

\subsection{Spatial evolution and the relaxation lengths} \label{sec:Lsat}

\begin{figure}
\begin{center}
\setlength{\unitlength}{1mm}
\begin{picture}(120, 52)(0, 46)
\put(2, 40){\includegraphics[width=12cm]{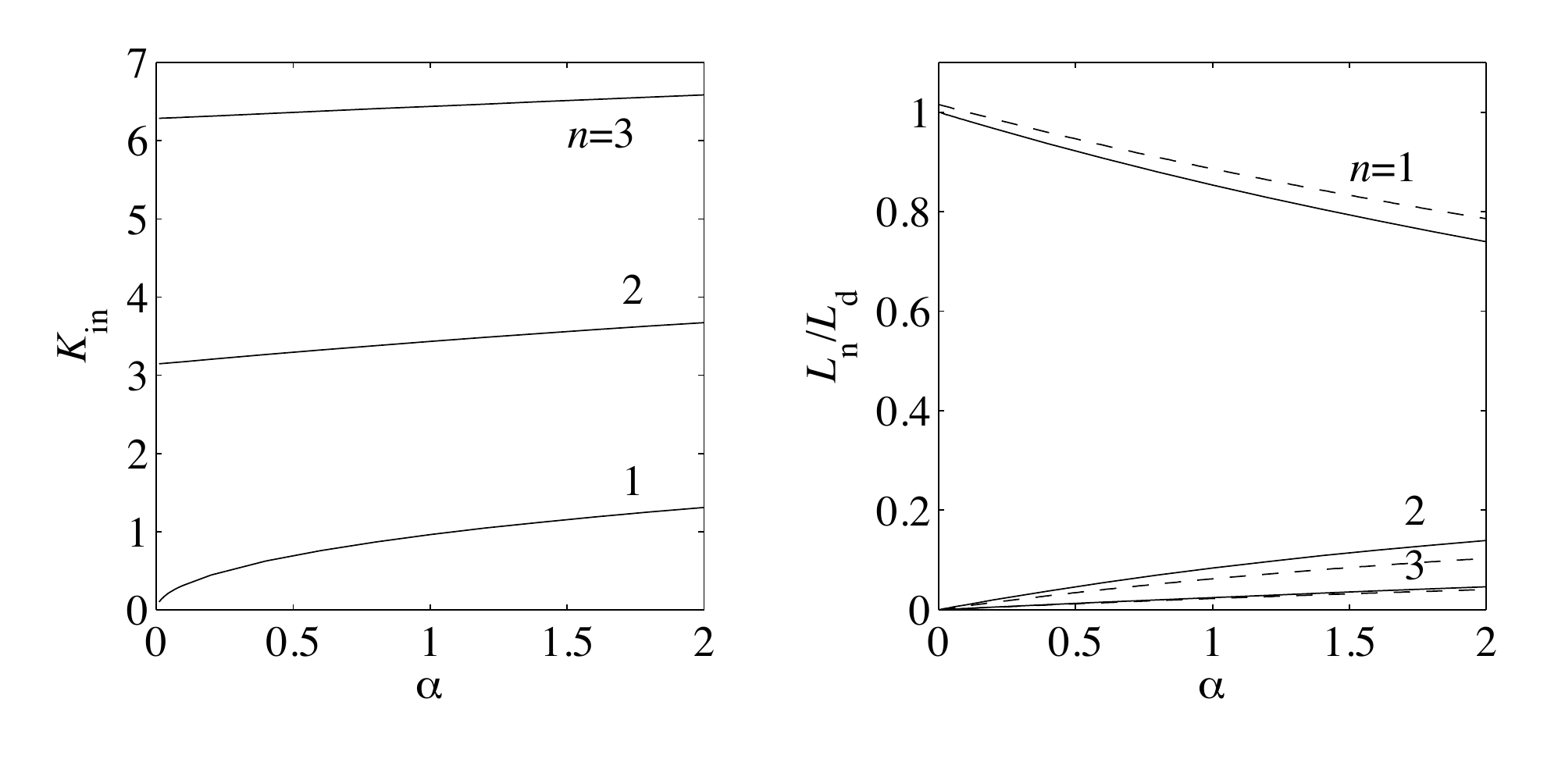}}
\put(3, 91){(a)}
\put(63, 91){(b)}
\end{picture}
\end{center}
\caption{(a) Variation with $\alpha$ of the three smallest roots of (\ref{eq:Ki}). (b) Corresponding relaxation normalised lengths $L_n/L_d$ for $\lambda/\mathcal{K} = 50$; plug flow (---); logarithmic flow ($--$).}
\label{fig:LnsurLd}
\end{figure}

We consider the situation where, for given flow conditions and corresponding saturated concentration profile $\Phi_{\rm sat}(z)$, the actual concentration profile at some point, say $x=0$, is $\Phi_{\rm sat}(z) - \phi(x=0,z)$ where $\phi(x,z)$ is a `concentration defect'. We search for the distance at which the saturated distribution $\Phi_{\rm sat}(z)$ is recovered, corresponding to vanishing $\phi(x,z)$. Looking for normal modes of relaxation of the concentration defect of the form 
\begin{equation}
\phi(x,z) = \Phi_{\rm b} f(z) \exp(-x/L),
\end{equation}
we get from the advection-diffusion equation (\ref{eq:concentration})
\begin{equation}
\left(\frac{\lambda u_*}{L}+\frac{D}{L^2}\right) f+\frac{d}{d z}\left(D \frac{d f}{d z}+f\,V_{\rm fall} \right) = 0.
\label{equamodale}
\end{equation}
At the free surface, the zero flux condition (\ref{eq:bc@z=H}) gives
\begin{equation}
D \frac{d f}{d z} + f\,V_{\rm fall} = 0 \qquad{\rm at}\quad z=H.
\label{eq:bc2@z=H}
\end{equation}
On the bed the friction velocity $u_*$ is assumed to be uniform. The erosion flux $\varphi_\uparrow$, which depends only on $u_*$, is uniform too. Hence, the disturbance of $\varphi_\uparrow$ is zero, so that, from (\ref{eq:bc@z=0}),
\begin{equation}
\frac{d f}{d z}=0 \qquad{\rm at} \quad z=0.
\label{eq:bc2@z=0}
\end{equation}
The above differential problem is solved numerically for parabolic $D$ (logarithmic flow model) and analytically for uniform $D=D_0$ (plug flow model). For uniform $D_0$, equation (\ref{equamodale}) has solutions of the form $f(z) \propto \exp (K z/H)$, where $K$ has to satisfy a quadratic equation with roots $K_+$ and $K_-$ given by
\begin{equation}
K_{\pm} = - \frac{\alpha}{2} \pm \mathrm{i} K_{\rm i}  \quad{\rm with}\quad 
K_{\rm i}=\sqrt{\frac{\lambda}{\mathcal{K}} \frac{H}{L} + \frac{H^2}{L^2} - \frac{\alpha^2}{4}}.
\label{eq:solK}
\end{equation}
Then the boundary conditions (\ref{eq:bc2@z=H}-\ref{eq:bc2@z=0}) select a discrete set of relaxation lengths $L$ verifying~:
\begin{equation}
\tan K_{\rm i} = \left( \frac{K_{\rm i}}{\alpha} - \frac{\alpha}{4K_{\rm i}} \right)^{-1}.
\label{eq:Ki}
\end{equation}
This equation has an infinite number of real positive solutions $K_{{\rm i}n}$, $n \ge 1$. Figure \ref{fig:LnsurLd}a shows the variation with $\alpha$ of the three smallest ones ($n = 1, 2, 3$). For small $\alpha$, these solutions behave as $K_{\rm i1} \sim \sqrt{\alpha}$ and $K_{{\rm i}n} \sim (n-1)\pi$ for $n\ge2$.

The corresponding relaxation lengths $L_n$ are found from (\ref{eq:solK}):
\begin{equation}
\frac{H}{L_n} = \frac{1}{2} 
\left( -\frac{\lambda}{\mathcal{K}}  \pm \
\sqrt{\left(\frac{\lambda}{\mathcal{K}}\right)^2 + \alpha^2 + 4 K_{{\rm i}n}^2 } \right),  \qquad n \ge 1.
\label{eq:elln}
\end{equation}
They are displayed for $n = 1, 2, 3$ in Figure \ref{fig:LnsurLd}b as a function of $\alpha$ (solid lines), normalised with the characteristic deposition length 
\begin{equation}
L_d \equiv \frac{U}{V_{\rm fall}} \, H.
\end{equation}
It can be seen that $L_1$ is much larger than the higher-order relaxation lengths -- typically by one order of magnitude. Remarkably, in the limit of small $\alpha$ (large flow velocity or small settling velocity), the largest length $L_1$ tends to $L_d$, whereas higher-order lengths remain on the order of the flow depth $H$:
\begin{equation}
L_1 \sim L_d, \qquad 
L_n \sim \frac{\lambda / \mathcal{K}}{ (n-1)^2 \pi^2} H 
\quad {\rm for} \quad n \ge 2
\end{equation}
Figure~\ref{fig:LnsurLd}b also displays the normalized relaxation lengths obtained from the logarithmic flow model (dashed lines), from numerical integration of (\ref{equamodale}) with parabolic $D$. It can be seen that these lengths are close to those from the plug flow model, especially for the largest length $L_1$. Note that $L_n/L_d$ is weakly sensitive to the value of $\lambda/\mathcal{K}$: doubling this ratio does not bring any visible change, at least for $\alpha \le 2$.

\begin{figure}
\begin{center}
\setlength{\unitlength}{1mm}
\begin{picture}(120, 52)(0, 46)
\put(2, 40){\includegraphics[width=12cm]{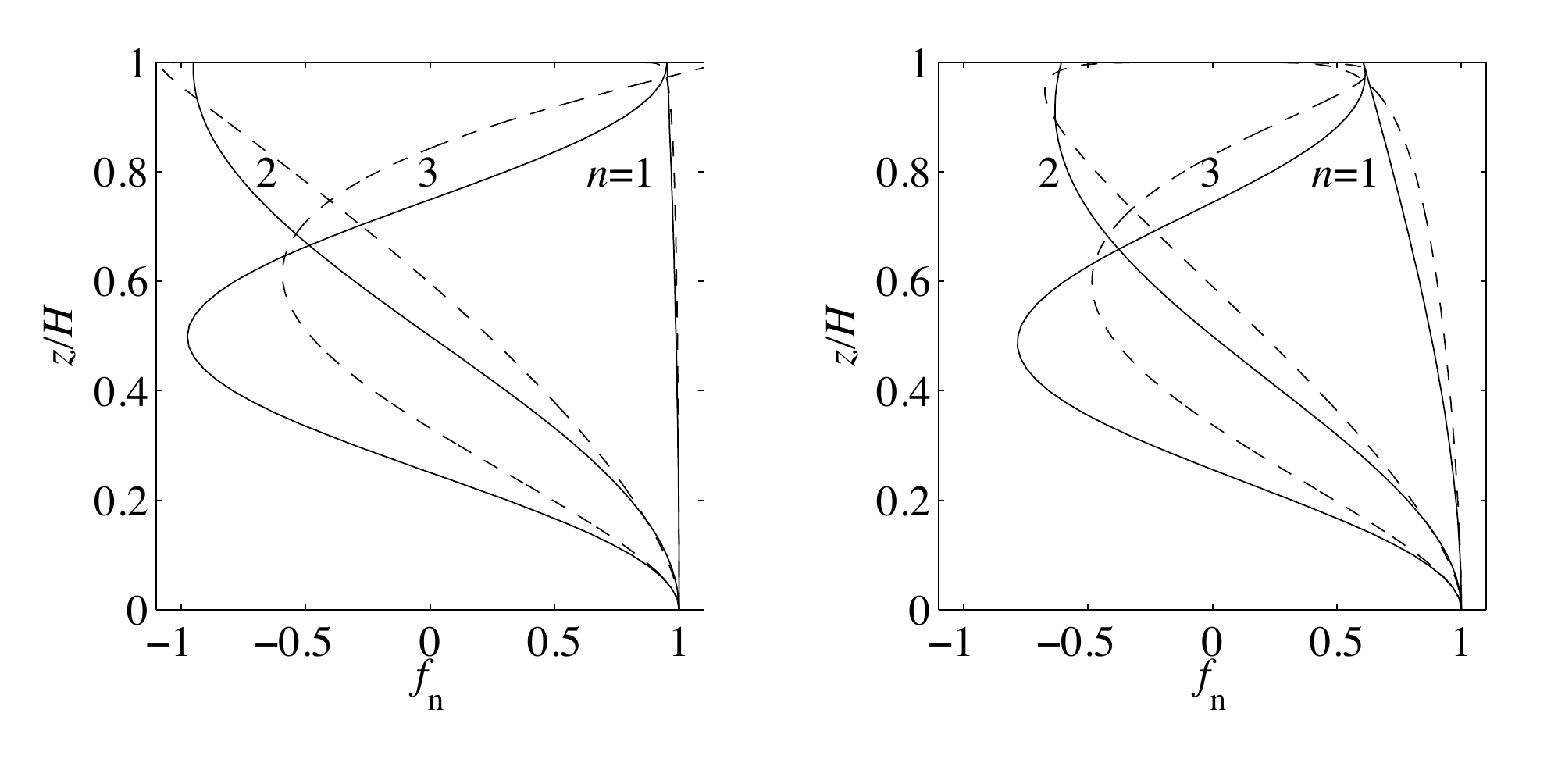}}
\put(3, 91){(a)}
\put(63, 91){(b)}
\end{picture}
\end{center}
\caption{Profile of the three first eigenfunctions $f_n(z)$, for $\alpha=0.1$ (a) and $\alpha=1$ (b). Plug flow (---); logarithmic flow ($--$).}
\label{fig:eigfun}
\end{figure}

The eigenfunctions $f_n(z)$ are given by
\begin{equation}
f_n(z) = \left[ \cos \left(K_{{\rm i}n}\frac{z}{H}\right) + \frac{\alpha}{2K_{{\rm i}n}} \, \sin \left(K_{{\rm i}n}\frac{z}{H}\right) \right] \exp\left( -\frac{\alpha}{2} \, \frac{z}{H} \right), \qquad n\ge1,
\label{eq:eigfun}
\end{equation}
with the normalization condition $f_n(0) = 1$. These eigenfunctions are displayed in Figure~\ref{fig:eigfun} for  $n = 1, 2, 3$ (solid lines), for $\alpha = 0.1$ (Fig.~\ref{fig:eigfun}a) and $\alpha = 1$ (Fig.~\ref{fig:eigfun}b). It can be seen that $f_1(z)$ decreases slightly and monotically from bottom to top, whereas higher-order eigenfunctions oscillate, more and more strongly with increasing $n$. Figure~\ref{fig:eigfun} also displays the eigenfunctions from the logarithmic flow model (dashed lines). It can be seen that for the mode associated with the largest length $L_1$ ($n=1$), eigenfunctions of both models remain very close to each other, and that differences become larger as $n$ increases. 

Let us turn to the sediment flux. The contribution of the $n^{\rm th}$-eigenmode to the sediment flux, $Q_n$, normalized with the characteristic sediment flux $U H \Phi_{\rm b}$ and the exponential $x$-dependence, is
\begin{equation}
\frac{1}{\exp(-x/L_n)} \, \frac{1}{\Phi_{\rm b} U H} \, Q_n= \frac{1}{H} \, \int_0^H f_n(z) {\rm d}z.
\label{eq:Qn}
\end{equation}
Table~\ref{tab:QnandAn}a displays the contribution of each of the first three modes to the sediment flux, \textit{i.e.} the right-hand side of the above equation. It can be seen that the contribution of the first mode $n=1$ strongly dominates. The smallness of the contribution of the higher-order modes is due to the oscillations of the eigenfunctions, as shown in Figure~\ref{fig:eigfun}. For small $\alpha$, the normalised flux is close to one for $n=1$ and decreases as $\alpha/((n-1)\pi)^2$ for $n \ge 2$.

\begin{table}
\begin{center}
(a) \ \ 
\begin{tabular}{|c|c|c|c|}
\hline
$n$ & 1 & 2 & 3 \\ \hline
$\alpha=0.1$ & 0.9836 & 0.0099 & 0.0025  \\
$\alpha=1$ & 0.8533 & 0.0832 & 0.0240 \\ \hline
\end{tabular}
\hspace*{1cm}
(b) \ \ 
\begin{tabular}{|l|c|c|c|c|}
\hline
            & $A_1$ & $A_2$ & $A_3$ & $A_4$ \\ \hline
$\alpha=0.1$ & 0.967 & 0.020 & 0.005 & 0.002 \\
$\alpha = 1$ & 0.724 & 0.150 & 0.047 & 0.022 \\ \hline
\end{tabular}
\caption{(a) Contribution of the three lowest-order eigenmodes to the normalized sediment flux (r.h.s. of equation (\ref{eq:Qn})), for $\alpha = 0.1$ and $\alpha = 1$. (b) Normalized coefficients $A_n = a_n / \delta \Phi_{\rm b}$ of the expansion (\ref{projection}), computed from a projection over four modes.}
\label{tab:QnandAn}
\end{center}
\end{table}

The general form of the concentration defect finally is
\begin{equation}
\phi(x,z) = \Phi_{\rm b} \, \sum_{n=1}^{\infty} a_n f_n(z)\exp(-x/L_n),
\label{eq:expansion}
\end{equation}
where the relaxation lengths $L_n$ are given by (\ref{eq:Ki}-\ref{eq:elln}), the eigenfunctions $f_n(z)$ are given by (\ref{eq:eigfun}), and the coefficients $a_n$ have to be determined by the concentration profile imposed at $x=0$. Such a determination will be illustrated in section \ref{sec:applic}.

\subsection{Temporal evolution and relaxation times} \label{sec:Tsat}

We now consider an  unsaturated concentration profile at initial time $t=0$, say $\Phi_{\rm sat}(z) - \phi(z, t=0)$, uniform in the streamwise $x$-direction, and search for the time needed for relaxation to the saturated distribution $\Phi_{\rm sat}(z)$ given by (\ref{eq:Phi}), \textit{i.e.} vanishing concentration defect $\phi(z, t)$. Calculations go along the same lines as in the previous sub-section, so they are only briefly sketched here. Looking for normal modes of the form
\begin{equation}
\phi(t,z) = \Phi_{\rm b} \, g(z) \exp(-t/T),
\end{equation}
we get from the equation (\ref{eq:concentration}) the equation governing the eigenfunctions $g(z)$:
\begin{equation}
\frac{1}{T} \, g+\frac{d}{d z}\left(D_0 \frac{d g}{d z}+g\,V_{\rm fall} \right)=0.
\label{equamodale4T}
\end{equation}
This equation has solutions of the form $g(z)\propto \exp(Kz/H)$, where $K$ has to satisfy a quadratic equation with roots $K_+$ and $K_-$ defined as
\begin{equation}
K_{\pm} = - \frac{\alpha}{2} \pm \mathrm{i} K_{\rm i}  \quad{\rm with}\quad 
K_{\rm i}=\sqrt{\frac{H}{\mathcal{K}u_*T} - \frac{\alpha^2}{4}}.
\label{eq:solK4T}
\end{equation}
The boundary conditions at $z=0$ and $z=H$ are the same as in the previous section, so that $K_{\rm i}$ still verifies equation (\ref{eq:Ki}), with same solutions $K_{{\rm i}n}$, $n \ge 1$. The corresponding relaxation times $T_n$ are then given by
\begin{equation}
\frac{H}{\mathcal{K}u_*T_n} = K_{{\rm i}n}^2 + \frac{\alpha^2}{4},  \qquad n \ge 1.
\label{eq:Tn}
\end{equation}
Introducing the characteristic deposition time 
\begin{equation}
T_d \equiv \frac{H}{V_{\rm fall}} = \frac{L_d}{U},
\end{equation}
the relaxation times are, in the limit of small $\alpha$, 
\begin{equation}
T_1 \sim T_d, \qquad 
T_n \sim \frac{\alpha T_d}{ (n-1)^2 \pi^2} \sim \frac{L_n}{U} 
\quad {\rm for} \quad n\ge2.
\end{equation}
As for the spatial problem, the sediment dynamics is dominated by the largest time $T_1$, equal to the deposition time $T_d$ for strong suspensions.

\section{Two illustrations, and comparison to experiments} \label{sec:applic}

\subsection{Effect of a change in the bed slope} \label{sec:slope}

Consider the situation depicted in Figure~\ref{fig:sketches}a, of a flow with saturated concentration profile $\Phi_0(z)$ which experiences a small variation $\delta S$ in the bottom slope at $x=0$, either positive or negative. This variation leads to a small change of the water depth and friction velocity, according to $\delta H/H = - \delta u_*/u_* = - \frac{1}{2} \delta S/S$. This change occurs on a hydrodynamic lengthscale $L_h$ given by the balance between the acceleration $U \delta U/L_h$ and the force $g \delta S$, \textit{i.e.} $L_h/L_d = U V_{\rm fall} / 2 u_*^2 = \lambda \mathcal{K} \alpha/2$. The present analysis is valid for small $L_h/L_d$, a condition which is fulfilled for small $\alpha$.

\begin{figure} 
\begin{center}
\setlength{\unitlength}{1mm}
\begin{picture}(120, 52)(0, 46)
\put(2, 40){\includegraphics[width=12cm]{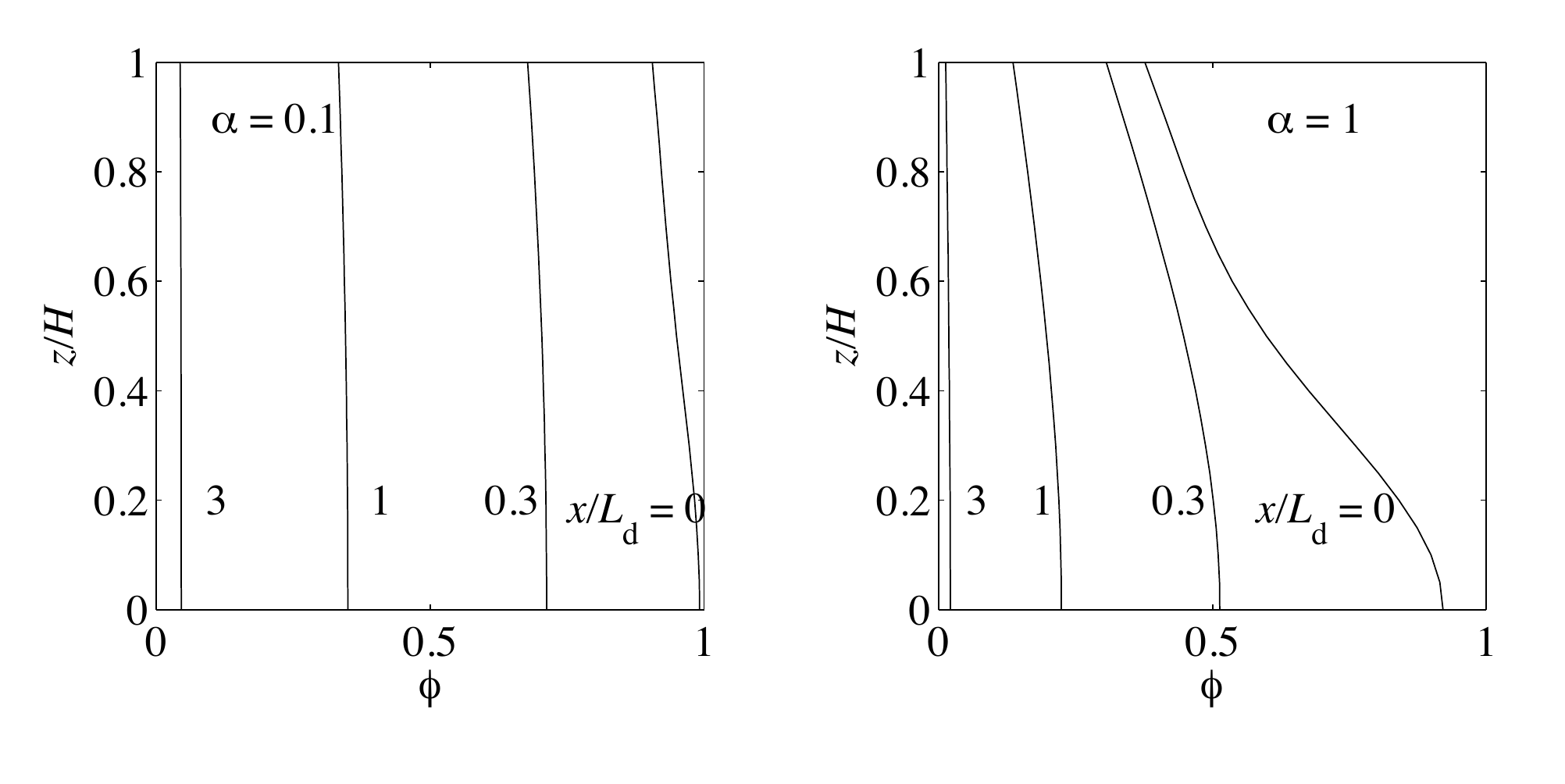}}
\put(3, 91){(a)}
\put(63, 91){(b)}
\end{picture}
\end{center}
\caption{Relaxation of the concentration defect $\phi(x,z)$ after a slope change as sketched in Figure~\ref{fig:sketches}a, with summation over the three first modes (plug flow model). (a), Profiles of $\phi(x,z)$ at the four downstream positions $x/L_d = 0$, 0.3, 1 and 3, for $\alpha = 0.1$; (b), same for $\alpha = 1$.}
\label{fig:phi_relax}
\end{figure}

The change in the saturated concentration profile due to the slope variation $\delta S$ is, at the linear order in $\delta u_*$,
\begin{equation}
\delta \Phi(z) =  \delta \Phi_{\rm b} 
             \exp \left( -\alpha \frac{z}{H} \right) \qquad \mbox{with} \qquad
 \delta \Phi_{\rm b} = \frac{\varphi'_\uparrow(u_*) \delta u_*}{V_{\rm fall}},
\label{PhiPhi}
\end{equation}
where $\varphi'_\uparrow(u_*)$ is the derivative of the erosion rate $\varphi_\uparrow(u_*)$. The concentration defect at $x=0$ corresponds to this change ($\phi(0,z) = \delta \Phi(z)$), so that the coefficients $a_n$ of the expansion (\ref{eq:expansion}) must satisfy

\begin{equation}
\delta \Phi_{\rm b} \exp \left( -\frac{\alpha}{2} \, \frac{z}{H} \right) = 
\sum_{n=1}^\infty a_n \left[ \cos \left(K_{{\rm i}n}\frac{z}{H}\right) + \frac{\alpha}{2K_{{\rm i}n}} \, \sin \left(K_{{\rm i}n}\frac{z}{H}\right) \right].
\label{projection}
\end{equation}
These coefficients can be determined from the projection of the above equation on the eigenfunctions, \textit{i.e.} truncation of the sum on the r.h.s. to $p$ terms, multiplication by each eigenfunction, and integration of both sides from $0$ to $H$. A linear system of $p$ equations is obtained, whose solution gives the coefficients $a_1$, ..., $a_p$. Table~\ref{tab:QnandAn}b displays the normalized coefficients $A_n = a_n/\delta \Phi_{\rm b}$ resulting from the projection over $p=4$ modes, for two values of $\alpha$. One can see that the first mode captures most of the weight; the contribution of the second one is smaller, but still significant, and higher modes are negligible. We have checked that considering more terms in the expansion has negligible effect on the dominant coefficients. The small weight of the oscillating modes is consistent with the slow variation with $z$ of the initial concentration profile (\ref{PhiPhi}).

Figure~\ref{fig:phi_relax} displays profiles of the concentration defect $\phi(x,z)$ at the location of the slope change, $x/L_d = 0$, and three positions downtream, for $\alpha = 0.1$ (Fig.~\ref{fig:phi_relax}a) and $\alpha = 1$ (Fig.~\ref{fig:phi_relax}b). It can be seen that at the position $x/L_d = 3$, the concentration defect is nearly zero. The flux defect, \textit{i.e.} the depth-integrated profiles of the concentration defect, correspondingly decays towards zero (not shown) and does so almost exponentially with relaxation length $L_d$, as predicted by equation (\ref{eq:relax}). This confirms that the dominant mode with relaxation length $L_d$ captures most of the sediment flux variations. For this example, as well as for the next ones, the question of the plug flow limit is not crucial: as far as the relaxation modes are concerned, the logarithmic and plug models do not differ much (Fig.~\ref{fig:eigfun}).

\subsection{Net erosion experiments} \label{sec:erodib}

Another situation of interest is that of a flow of clear fluid on a non-erodible bed ($\Phi_0(z)=0$) reaching an erodible bed lying in $x>0$, as sketched in Figure~\ref{fig:sketches}b. Suspension develops downstream until the saturated concentration profile $\Phi_{\rm sat}(z)$ is reached. The analysis goes along the same lines as for the slope change. The concentration defect $\phi(x,z)$, can be decomposed on the eigenfunctions (\ref{eq:eigfun}), with coefficients $a_n$ determined by the concentration profile at $x=0$. The equation to be satisfied turns out to be the same as (\ref{projection}) with $\Phi_{\rm b}$ instead of $\delta \Phi_{\rm b}$. Thus the coefficients are $a_n = A_n \delta \Phi_{\rm b}$ with the normalised coefficients $A_n$ given in Table~\ref{tab:QnandAn}b.

\begin{table}
\begin{center}
\begin{tabular}{|l|c|c|c|c|}
\hline
 & van Rijn (1986b) 
 & \multicolumn{2}{c|}{Ashida \& Okabe (1982)} 
 & Jobson \& Sayre (1970) \\ 
\hline
 & & \quad run 5 \quad \quad & run 6 & runs FS1 and FS1A  \\
 & erosion & erosion  & deposition  & deposition  \\
Symbol & ({\Large $\ast$}) & ({\Large $\circ$}) 
       & ($\square$) & ($\blacksquare$) \\
$H$ (cm)                & 25   & \multicolumn{2}{c|}{4.3}  & 40.7  \\ 
$U$ (cm/s)              & 67   & \multicolumn{2}{c|}{37.3} & 29.1  \\
$u_*$ (cm/s)            & 4.77 & \multicolumn{2}{c|}{3.63} & 4.48  \\
$V_{\rm fall}$ (cm/s)   & 2.2  & \multicolumn{2}{c|}{1.85} & 1.0  -- 2.0  \\
$L_d$ (m)               & 7.6  & \multicolumn{2}{c|}{0.87} & 5.9  -- 11.9 \\
$\alpha$                & ---  & 3.1  & 2.5                & 1.2          \\
Sc                      & ---  & 0.41 & 0.33               & 0.18 -- 0.36 \\
\hline
\end{tabular}
\end{center}
\caption{Hydraulic parameters $H$, $U$, $u_*$ and $V_{\rm fall}$ of the experiments, and $L_d =  (U/V_{\rm fall})H$, $\alpha$ from the exponential fit of the downstream concentration profile, and ${\rm Sc} = \kappa \alpha u_*/6 V_{\rm fall}$.}
\label{tab:param_exp}
\end{table}

\begin{figure} 
\begin{center}
\setlength{\unitlength}{1mm}
\begin{picture}(120, 60) 
\put(0, 0){\includegraphics[width=12cm]{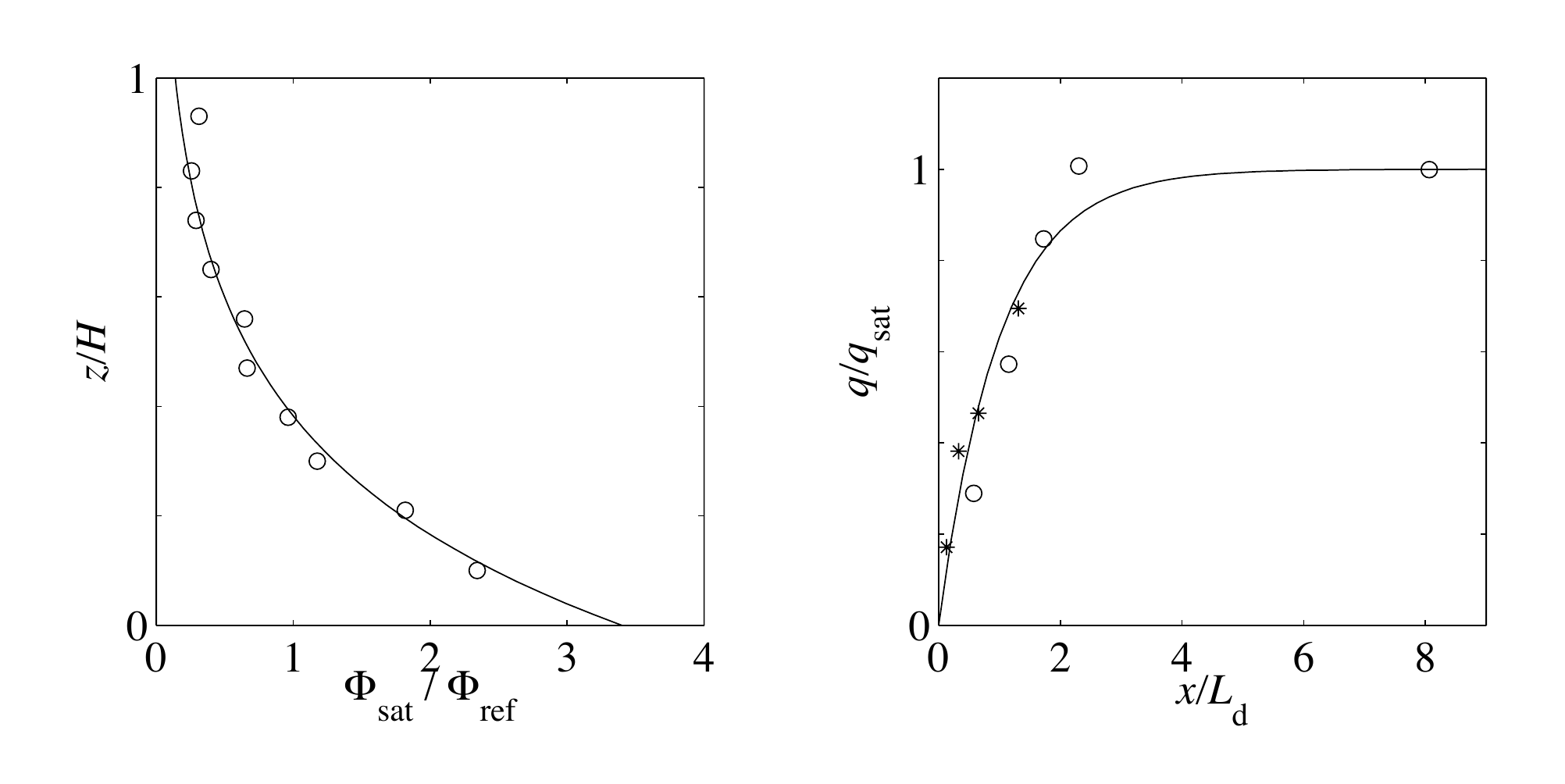}}
\put(47, 49){(a)}
\put(107, 49){(b)}
\end{picture}
\end{center}
\vspace*{0.2cm}
\caption{Net erosion experiments. (a) Concentration profile at the farthest location $x/L_d = 8.1$ measured by Ashida \& Okabe (1982, run 5), normalized with the depth-averaged concentration $\Phi_{\rm ref}$; solid line: $\Phi_{\rm sat}$ given by (\ref{eq:Phi}) with $\alpha = 3.2$. (b) Relaxation to saturation of the normalized sediment flux versus $x/L_d$; symbols: van Rijn (1986b) and Ashida \& Okabe (1982) (see Table~\ref{tab:param_exp}); solid line: exponential relaxation (\ref{eq:relax_eros}).}
\label{fig:relax1}
\end{figure}

The prediction that the eigenmode with the largest relaxation length captures most of the sediment flux can be assessed from the experimental observations of van Rijn (1986b) and Ashida \& Okabe (1982) --  non-Japanese readers can access these latter data in the paper of Celik \& Rodi (1988). These experiments precisely correspond to the sketch depicted in Figure~\ref{fig:sketches}b. Their hydraulic parameters are given in Table~\ref{tab:param_exp}. The spatial evolution of the concentration profiles has been measured at different locations downstream the transition point at $x=0$. The corresponding sediment flux $q$, which is zero for $x<0$, increases downstream until it reaches the saturated value $q_{\rm sat}$. We determined this flux from integration of the measured concentration profile at each $x$-location. From the hydraulic parameters, the deposition length can be computed as $L_d = (U/V_{\rm fall}) H$ -- in the following we will not distinguish between $L_1$ and $L_d$, although they can differ by $\approx 20 \%$ for $\alpha$ on the order of unity (Fig.~\ref{fig:LnsurLd}b). It appeared that for Ashida \& Okabe (1982) the location of the farthest downstream measurements corresponds to $x/L_d = 8.1$, which is large enough for the sediment flux to be saturated. Figure~\ref{fig:relax1}a displays the corresponding concentration profile, and an exponential fit providing, from  (\ref{eq:Phi}), the value of the parameter $\alpha$ reported in Table~\ref{tab:param_exp}. For van Rijn (1986b), we found  $x/L_d = 1.3$, not large, preventing any straightforward determination of the parameter $\alpha$. For both experiments, the saturated flux $q_{\rm sat}$ was estimated as that providing the best fit to the exponential curve 
\begin{equation}
\frac{q}{q_{\rm sat}} = 1 - \exp(-x/L_d).
\label{eq:relax_eros}
\end{equation}
Figure~\ref{fig:relax1}b diplays the variation of $q/q_{\rm sat}$ with $x/L_d$; it can be seen that the data points fall quite well on the exponential curve. Note that the deposition lengths $L_d$ differ by one order of magnitude between the two experiments. The data collapse therefore supports a first-order relaxation process with characteristic length equal to $L_d$.

\subsection{Net deposition experiments} \label{sec:depos}

\begin{figure} 
\begin{center}
\setlength{\unitlength}{1mm}
\begin{picture}(120, 60) 
\put(0, 0){\includegraphics[width=12cm]{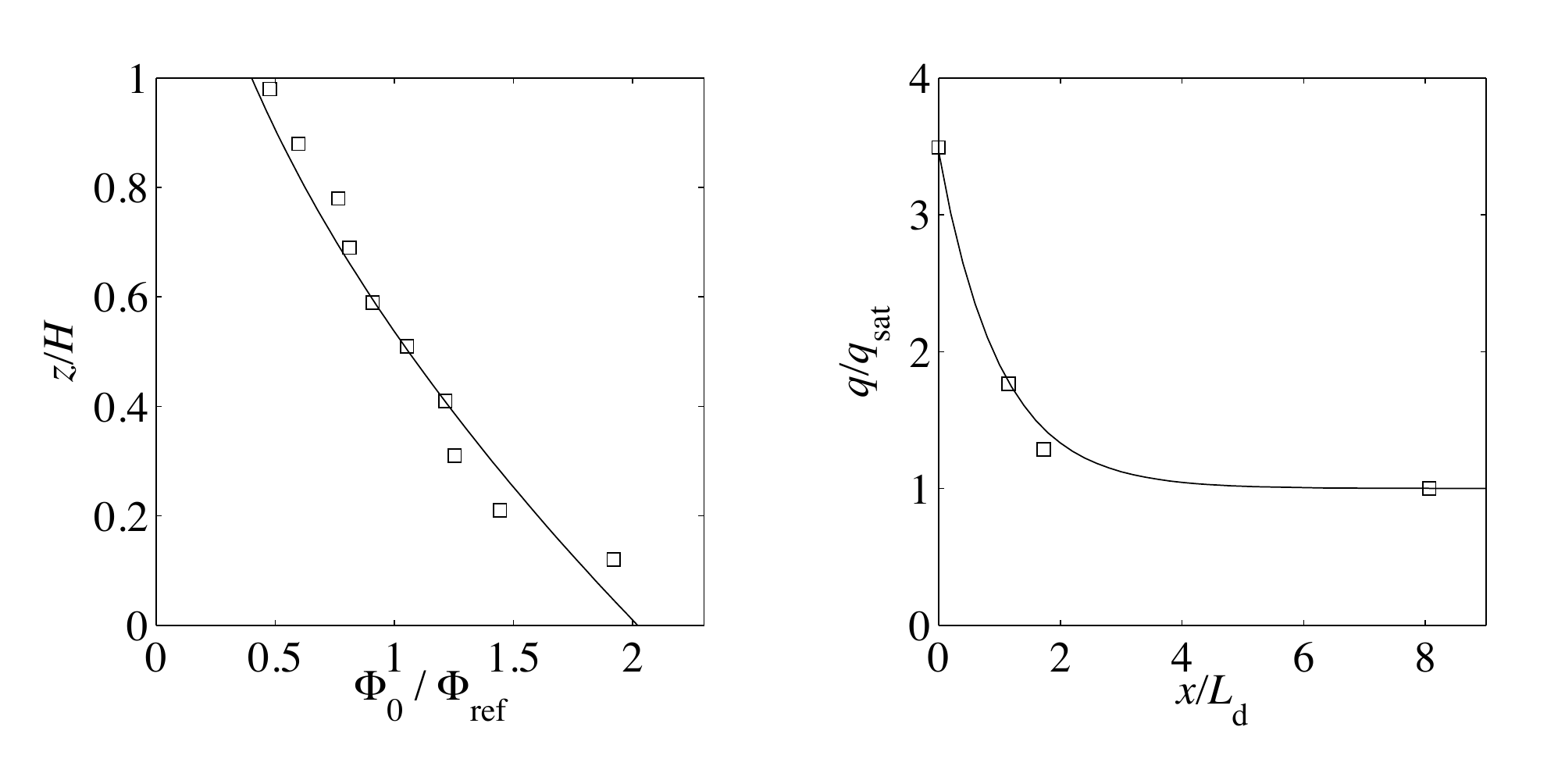}}
\put(47, 49){(a)}
\put(107, 49){(b)}
\end{picture}
\end{center}
\vspace*{0.2cm}
\caption{Net deposition experiments of Ashida \& Okabe (1982) (run 6, see Table \ref{tab:param_exp}). (a) Concentration profiles at $x=0$ normalized with the depth-averaged concentration $\Phi_{\rm ref}$; solid line: $\Phi_0$ reconstructed from $\Phi_{\rm sat}$ and one single mode for the concentration defect.
(b) Normalized sediment flux versus $x/L_d$, experiments and exponential relaxation (\ref{eq:relax_depos}).}
\label{fig:relax2}
\end{figure}

Ashida \& Okabe (1982) have also performed experiments in which the initial concentration profile is oversaturated (run 6), \textit{i.e.} the initial sediment flux $q_0$ at $x=0$ is larger than $q_{\rm sat}$, so that the sediment settle until the saturated regime is reached further downstream. Figure~\ref{fig:relax2}b displays the sediment flux, obtained from the measured concentration profiles, together with the exponential relaxation curve now given by
\begin{equation}
\frac{q}{q_{\rm sat}} = 
     1 + \left( \frac{q_0}{q_{\rm sat}} - 1 \right) \exp(-x/L_d)
\label{eq:relax_depos}
\end{equation}
where $q_0$ and $q_{\rm sat}$ were determined by curve fitting, and $L_d$ is given in Table~\ref{tab:param_exp}. Again, the agreement is quite good, showing that the mode with relaxation length $L_d$ captures most of the deposition process.

As a confirmation, Figure~\ref{fig:relax2}a compares the concentration profile measured at the location $x=0$ to its projection over one single mode. This projection is computed from an empirical representation of this inital profile,  using the expansion (\ref{eq:expansion}) and the saturated flux $\Phi_{\rm sat}(z)$ measured from the concentration profile at $x = 8.1\,L_d$, fitted by the exponential form (\ref{eq:Phi}) with $\alpha = 2.5$. It can be seen that the resulting profile is in good agreement with the measurements. Note that $\alpha = 2.5$ corresponds to Schmidt number ${\rm Sc} = \alpha \kappa u_*/6 V_{\rm fall} = 0.33$, which is slightly below the usual range $0.5$--$1$ (Coleman 1970; Celik \& Rodi 1988; Nielsen 1992).

\begin{figure} 
\begin{center}
\setlength{\unitlength}{1mm}
\begin{picture}(120, 60) 
\put(0, 0){\includegraphics[width=12cm]{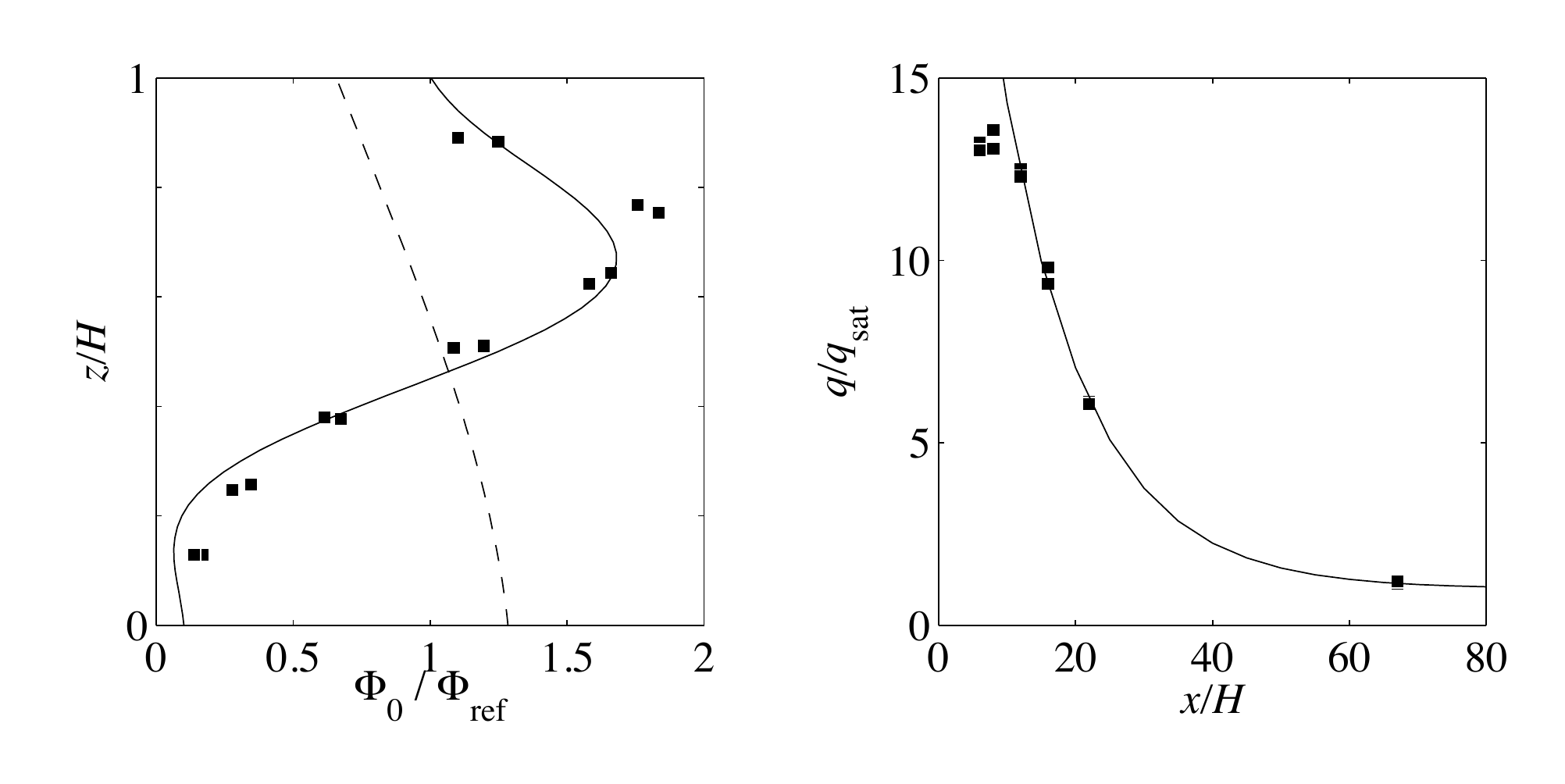}}
\put(47, 49){(a)}
\put(107, 49){(b)}
\end{picture}
\end{center}
\vspace*{0.2cm}
\caption{Net deposition experiments of Jobson \& Sayre (1970) (see Table \ref{tab:param_exp}). (a) Measured concentration profile at $x=0$ normalized with the depth-averaged concentration $\Phi_{\rm ref}$; dashed and solid lines: $\Phi_0$ reconstructed with one single mode and four modes, respectively, and $\alpha = 1.2$.
(b) Relaxation to saturation of the normalized sediment flux versus $x/L_d$; solid line: exponential relaxation (\ref{eq:relax_depos}) with $L_d = 5.2$ m.}
\label{fig:relax3}
\end{figure}

Other net deposition experiments have been performed by Jobson \& Sayre (1970). In this work, the particles were released near the water surface, so that the initial concentration profiles exhibits a peak close to $z=H$, as shown in Figure~\ref{fig:relax3}a. In contrast to Ashida \& Okabe experiments, expanding the concentration defect over one single mode is not sufficient to get a good representation of this profile; an expansion over four modes provide a much better description, as shown in Figure~\ref{fig:relax3}a. However, the high-order modes are expected to vanish over a short distance, on the order of a few flow depths $H$, and the exponential relaxation to be recovered at large distances. This scenario is evidenced in Figure~\ref{fig:relax3}b, which displays the normalized flux versus $x/H$, measurements and the exponential curve (\ref{eq:relax_depos}). Here, due to uncertainties on the falling velocity (see Table~\ref{tab:param_exp} of the present paper and Figure~6a of Jobson \& Sayre 1970), the deposition length $L_d$ was determined, together with $q_0$ and $q_{\rm sat}$, by fitting the experimental data points. We found $L_d = 5.2$ m, which is close to the range of the expected values displayed in Table~\ref{tab:param_exp}, although slightly smaller. We finally note that in the course of the reconstruction of $\Phi_0(z)$, we found  $\alpha = 1.2$ from the saturated concentration profile, which corresponds to Schmidt number in the range 0.18--0.36, slightly smaller, again, than the usual range.

\section{Concluding remarks} \label{sec:conclusion}

In this paper, we have discussed the conditions under which a first-order relaxation equation for the sediment flux $q$ can be derived for turbulent flows, when suspension is the dominant mode of transport. From a mode analysis of the linear advection-diffusion equation for the particle concentration, it was shown that the sediment flux is dominated by the mode corresponding to the largest relaxation length for spatially varying flows, or the largest relaxation time for time-dependent flows. These relaxation scales were identified as the deposition length $H U/V_{\rm fall}$ and the deposition time $H/V_{\rm fall}$, where $H$ is the flow depth, $U$ the mean flow velocity and $V_{\rm fall}$ the sediment settling velocity. This result is expected to be particularly relevant for the case of sediment transport in slowly varying flows, for which the flux is never far from saturation. Predictions of the sediment flux were shown to be in quantitative agreement with flume experiments, for both net erosion and net deposition situations, and deposition lengths spanning over one order of magnitude.

As discussed in the introduction, the relaxation equation~(\ref{eq:relax}) allows for the description of both bed load and suspended load. However, these modes of transport correspond to very different physical lengthscales. For bed load, the relaxation length $L_{\rm sat}$ is on the order of $10$ grain diameters (Fourri\`ere, Claudin \& Andreotti 2010). As soon as the flow depth $H$ is larger than a few $L_{\rm sat}$, the -- unstable -- flat bed is insensitive to the presence of the free surface, and current ripples emerge at a centimetric wavelength ($\approx 10 L_{\rm sat}$). When suspended load is the dominant type of transport, we have shown that the relaxation length  $L_{\rm sat}$ is on the order of $10$--$100~H$, which is typically four to five orders of magnitude larger than for bed load. Suspended transport thus prevents the formation of bedforms with wavelength smaller than $H$, and patterns such as bars, antidunes and meanders can be expected to emerge from linear instability, with large wavelengths on the order of $100$--$1000~H$. Further work is required for the experimental and theoretical investigations of these instabilities.

\vspace*{0.3cm}

\noindent
\rule[0.1cm]{3cm}{1pt}

\noindent
This work has benefited from the financial support of the Agence Nationale de la Recherche, grant `Zephyr' ($\#$ERCS07\underline{\ }18) and the GdR `M\'ePhy' of the CNRS ($\#$3166).


\end{document}